\documentclass[argument]{aastex62}
\usepackage{amssymb}
\usepackage{amsmath}
\usepackage{gensymb}
\usepackage{graphicx}
\usepackage{xcolor}
\usepackage{natbib}

\newcommand{\hi}{\ifmmode{\rm HI}\else{H\/{\sc i}}\fi}
\newcommand{\HI}{H\,{\sc i} }
\newcommand{\msol}{\mbox{$M_\odot$} }

\newcommand {\kms}{\ifmmode{\rm km \, s^{-1}}\else{$\rm km \, s^{-1}$}\fi}

\shorttitle{Formation of a massive lenticular galaxy}
\shortauthors{Jin-long Xu et al.,}
\begin{document}

%	\title{Half of the baryonic mass in star-forming galaxies at $z\approx1.3$ is Atomic Hydrogen}
	
	\title{Formation of a massive lenticular galaxy under the tidal interaction with a group of dwarf galaxies}	

\correspondingauthor{Jin-Long Xu}
\email{xujl@bao.ac.cn}

\author{Jin-Long Xu}
\affiliation{National Astronomical Observatories, Chinese Academy of Sciences, Beijing 100101, People's Republic of China}
\affil{Guizhou Radio Astronomical Observatory, Guizhou University, Guiyang 550000, People's Republic of China}
\affil{CAS Key Laboratory of FAST, National Astronomical Observatories, Chinese Academy of Sciences, Beijing 100101, People's Republic of China}

\author{Ming Zhu}
\affiliation{National Astronomical Observatories, Chinese Academy of Sciences, Beijing 100101, People's Republic of China}
\affil{Guizhou Radio Astronomical Observatory, Guizhou University, Guiyang 550000, People's Republic of China}
\affil{CAS Key Laboratory of FAST, National Astronomical Observatories, Chinese Academy of Sciences, Beijing 100101, People's Republic of China}

\author{Kelley M. Hess}
\affiliation{ASTRON-Netherlands Institute for Radio Astronomy, Postbus 2, 7990 AA Dwingeloo, the Netherlands}
\affiliation{Kapteyn Astronomical Institute, University of Groningen Postbus 800, 9700 AV Groningen, the Netherlands}

\author{Naiping Yu}
\affiliation{National Astronomical Observatories, Chinese Academy of Sciences, Beijing 100101, People's Republic of China}
\affil{Guizhou Radio Astronomical Observatory, Guizhou University, Guiyang 550000, People's Republic of China}
\affil{CAS Key Laboratory of FAST, National Astronomical Observatories, Chinese Academy of Sciences, Beijing 100101, People's Republic of China}

\author{Chuan-Peng Zhang}
\affiliation{National Astronomical Observatories, Chinese Academy of Sciences, Beijing 100101, People's Republic of China}
\affil{Guizhou Radio Astronomical Observatory, Guizhou University, Guiyang 550000, People's Republic of China}
\affil{CAS Key Laboratory of FAST, National Astronomical Observatories, Chinese Academy of Sciences, Beijing 100101, People's Republic of China}

\author{Xiao-Lan Liu}
\affiliation{National Astronomical Observatories, Chinese Academy of Sciences, Beijing 100101, People's Republic of China}
\affil{Guizhou Radio Astronomical Observatory, Guizhou University, Guiyang 550000, People's Republic of China}
\affil{CAS Key Laboratory of FAST, National Astronomical Observatories, Chinese Academy of Sciences, Beijing 100101, People's Republic of China}

\author{Mei Ai}
\affiliation{National Astronomical Observatories, Chinese Academy of Sciences, Beijing 100101, People's Republic of China}
\affil{Guizhou Radio Astronomical Observatory, Guizhou University, Guiyang 550000, People's Republic of China}
\affil{CAS Key Laboratory of FAST, National Astronomical Observatories, Chinese Academy of Sciences, Beijing 100101, People's Republic of China}

\author{Peng Jiang}
\affiliation{National Astronomical Observatories, Chinese Academy of Sciences, Beijing 100101, People's Republic of China}
\affil{Guizhou Radio Astronomical Observatory, Guizhou University, Guiyang 550000, People's Republic of China}
\affil{CAS Key Laboratory of FAST, National Astronomical Observatories, Chinese Academy of Sciences, Beijing 100101, People's Republic of China}

\author{Jie Wang}
\affiliation{National Astronomical Observatories, Chinese Academy of Sciences, Beijing 100101, People's Republic of China}

	
	
	\begin{abstract}
     Based on the  atomic-hydrogen (\hi) observations using the Five-hundred-meter Aperture Spherical radio Telescope (FAST), we present a detailed study of the gas-rich massive S0 galaxy NGC 1023 in a nearby galaxy group.  The presence of an \HI extended warped disk in NGC 1023 indicates that this S0 galaxy originated from a spiral galaxy. The data also suggest that NGC 1023 is interacting with four dwarf galaxies. In particular, one of the largest dwarf galaxies has fallen into the gas disk of NGC 1023, forming a rare bright-dark galaxy pair with a large gas clump. This clump shows the signature of a galaxy but has no optical counterpart, implying that it is a newly formed starless galaxy. Our results firstly suggest that a massive S0 galaxy in a galaxy group can form via the morphological transformation from a spiral under the joint action of multiple tidal interactions.	\end{abstract}
	
	\keywords{galaxies: dwarf -- galaxies: evolution -- galaxies: formation}
	
\section{Introduction}

One of the biggest challenges for extragalactic astrophysics is to figure out how different types of galaxies are formed. Lenticular (S0) galaxies, located between ellipticals and spirals on the Hubble sequence, are composed of a central bulge  and a stellar disc without spiral arms. Two main formation pathways for S0 galaxies have been identified, either as fading spirals or as the result of galaxy mergers \citep{Fraser-McKelvie+18,Deeley+20,Deeley+21}. However, recent observations have demonstrated that S0 galaxies form a diverse population \citep{Coccato+20}, so there are other plausible mechanisms for their formation that still need to be probed.

It has been suggested that S0 galaxies descended from spirals through morphological transformation \citep{Dressler+80,Laurikainen+10,Rizzo+18}. Some dynamic processes may deplete the gas of the spirals and stop their star formation \citep{Larson+80}. Different processes would have different effects on the stellar and gas kinematics in S0 galaxies. The stellar kinematics have mostly been probed using  spatially resolved stellar or globular cluster distributions \citep{Herrmann+09,Zanatta+18}. Simulations indicate that the stellar kinematics have a velocity dispersion as high or even higher than  rotational velocity, if the S0 galaxies are the result of merger events \citep{Bournaud+05}. In the scenario of a fading spiral, the gas in the spiral arms is stripped by environmental mechanisms such as ram-pressure stripping, harassment, or  starvation \citep{Gunn+72,Moore+98,Balogh+00}. These relatively gentle gas-stripping processes should not have a significant impact on the stellar kinematics of the spirals, which would lead to a rotation that  would dominate the disks of the S0 galaxies \citep{Coccato+20,Deeley+20}. 

The NGC 1023 group is one of the most extensively studied nearby galaxy groups. Optical observations towards the group have detected dozens of dwarf galaxies \citep{Trentham+09}. The massive S0 galaxy NGC 1023 is the brightest member of this group at a distance of 10.4 Mpc \citep{Cappellari+11,Tonry+01,Dolfi+21}. The gas content of early-type galaxies is often thought to be insignificant. NGC 1023 is  one of the nearest members of the rare class of \hi-rich S0 galaxies, making it an ideal laboratory to test a massive S0 galaxy formation within a galaxy group. NGC 1023 has a complex set of substructures, including a young (3.4 Gyr) and metal-rich ([Fe/H]= 0.50 dex) nuclear disc  and a faster bar \citep{Corsini+16,Debattista+02}. Deep \HI observations from the Westerbork Sybthesis Radio Telescope (WSRT) showed that NGC 1023 has an extended and complex gas distribution \citep{Sancisi+84,Morganti+06}. Within the innermost part of the NGC 1023 group, there are four late-type  dwarf companion galaxies \citep{Trentham+09}, called NGC 1023A, NGC 1023B, NGC 1023C, and NGC 1023D. Only NGC 1023A was considered to be an ongoing interaction with the main galaxy NGC 1023  and that it has been stripped of its gas \citep{Capaccioli+86,Dolfi+21,Silva+22}, resulting in an extended and irregular \HI distribution detected in NGC 1023.

The kinematics of stars, globular clusters (GCs) and planetary nebulae (PNe) in NGC 1023 imply that  
its stellar disc is rotationally supported at small radii ($V_{\rm rot}/\sigma>1$), while more pressure supported at large radii ($V_{\rm rot}/\sigma<1$) \citep{Noordermeer+08,Dolfi+21}. Here $V_{\rm rot}$ and $\sigma$ represent rotation velocity and velocity dispersion. Furthermore, the peak $V_{\rm rot}/\sigma$ in its stellar disc is greater than 4 \citep{Noordermeer+08,Cortesi+11,Cortesi+13,Cortesi+16}, which is  higher than that ($1<V_{\rm rot}/\sigma\leq2$) of minor mergers ($3<$ mass ratio $<10$) and that ($V_{\rm rot}/\sigma<1$) of major mergers (mass ratio $<3$) \citep{Bournaud+05}, but  lower than the one expected for spiral galaxies \citep{Herrmann+09}. Both the fading spiral and merger scenarios appear to be inconsistent with the observed stellar kinematics of NGC 1023. Moreover, considering a lower luminosity of NGC 1023A relative to NGC 1023 \citep{Capaccioli+86,Noordermeer+08}, NGC 1023A is not sufficiently massive to cause a strong morphological response in NGC 1023.  \HI gas emission is typically more extended than stars in the outer parts of the galaxy. \HI observations can be considered as an excellent tool to investigate S0 galaxy formation. 

\subsection{Observations and data reduction.}
Using the Five-hundred-meter Aperture Spherical radio Telescope (FAST) on December 2021 \citep{Jiang+19,Jiang+20}, we have
performed a 2$^{\circ}\times$ 2$^{\circ}$ mapping observation of \HI (1420.4058 MHz) towards S0 galaxy NGC 1023.  The mapping observation used the Multi-beam on-the-fly (OTF) mode. This mode is designed to map the sky with 19 beams simultaneously, and  has a similar scanning trajectory. Besides, we set the scan velocity as 15$^{\prime\prime}$ s$^{-1}$ and an integration time of 1 second per spectrum.   It formally works in the frequency range from 1050 MHz to 1450 MHz.  We used the digital Spec(W) backend, which has a bandwidth of 500 MHz and 64k channels, resulting in a frequency resolution of 7.629 kHz and corresponding to a  velocity resolution of 1.6 \kms. For intensity calibration to antenna temperature ($T_{\rm A}$), noise signal with amplitude of 10 K was injected in a period of 32 seconds. The half-power beam width (HPBW) is $\sim$2.9$^{\prime}$ at 1.4 GHz for each beam. The pointing accuracy of the telescope was better than 10$^{\prime\prime}$.  To improve sensitivity, we observed two times for the NGC 1023 region, using a total of 2.3 hr. System temperature was around 25 K during the observations. The detailed data reduction is similar to \citet{Xu2021}.  Using the ArPLS algorithm \citep{Baek+15}, we mitigated radio frequency interference (RFI) by a fitting procedure of the data in the time-frequency domain. Finally,  we made the calibrated data into the standard cube data, with a pixel of $1.0^{\prime}\times1.0^{\prime}$. A gain $T_{\rm A}/\it S_{v}$ has been measured to be about 15 K Jy$^{-1}$. While a measured relevant main beam gain $T_{\rm B}/\it S_{v}$ is about 20 K Jy$^{-1}$ at 1.4 GHz for each beam, where $T_{\rm B}$ is the brightness temperature.  To construct a highly-sensitive \HI image, the velocity resolution of the FAST data  is smoothed to $\sim$3.2 $\kms$. The mean noise RMS in the observed image is $\sim$1.3 mJy beam$^{-1}$ in the observed region.

In addition, Digitized Sky Survey (DSS) Blue-band and GALEX NUV and FUV data are available at NASA/IPAC Extragalactic Database \footnote{http://ned.ipac.caltech.edu/}. The Hubble Space Telescope (HST) data are publicly available at Hubble Legacy archive\footnote{https://hla.stsci.edu/}.

\begin{figure}
\centering
\includegraphics[width=0.98\textwidth]{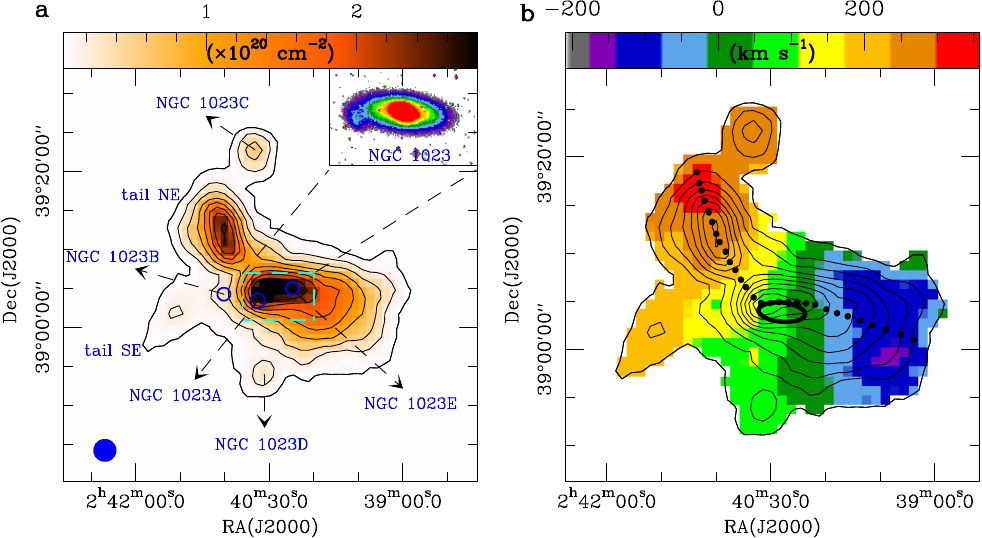}
\caption{Overview of the new \HI observations of NGC 1023.  {\bf a}, \HI column-density map shown in colour scale and black contours. The black contours begin at 4.0$\times$10$^{18}$ cm$^{-2}$ (3$\sigma$) in steps of 3.9$\times$10$^{19}$ cm$^{-2}$. The Digitized Sky Survey (DSS) B-band image of NGC 1023 is shown in the top-right corner, while the FAST beam in a green filled circle is shown in the bottom-left corner. {\bf b}, \HI velocity field (Moment 1) map in colour scale overlaid with  the \HI column-density in the black contours. These velocities are relative to the systemic velocity (615 \kms) of NGC 1023. The black ellipse indicates the position and size of NGC 1023. We use a dotted line to indicate velocity gradient.  This dotted line roughly follows the main axis of \HI complex of NGC 1023.}
\label{fig:NGC1023_all}
\end{figure}

\begin{figure}
\centering
\includegraphics[width=0.97\textwidth]{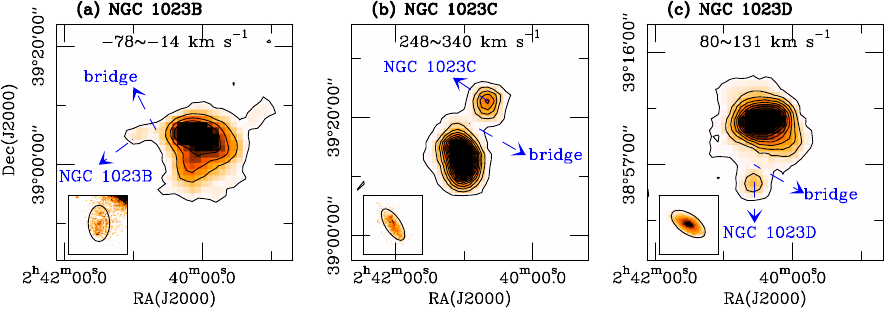}
\caption{{\noindent  Three dwarf galaxies interacting with NGC 1023}. The \HI column-density map of each dwarf galaxy is shown in colour scale and black contours, while the Pan-STARRS1 g-band image is shown in the lower-left corner of each panel.   The \HI column density in each panel was created within a certain integrated-velocity  range around each dwarf galaxy.  Relative to the systemic velocity (615 \kms) of NGC 1023, the integrated-velocity range for each dwarf galaxy is also presented in the top of each panel, while the black contours begin at 3.3$\times$10$^{18}$ cm$^{-2}$ in steps of 1.2$\times$10$^{19}$ cm$^{-2}$ in each panel.}
\label{fig:NGC1023_group}
\end{figure}

\section{Results}
We have performed deeper mapping observations of the \HI emission towards NGC 1023 using the Five-hundred-meter Aperture Spherical radio Telescope (FAST). The final data cube had a rms noise level of 1.3 mJy beam$^{-1}$ in the line channels, corresponding to a column density of 1.5$\times$10$^{17}$ cm$^{-2}$ channel$^{-1}$ at a velocity resolution of $\sim$3.2 \kms. We found that the \HI gas emission associated with NGC 1023 is mainly located within the redshift velocity range of 406.0 $\kms$ to 972.0 $\kms$. Relative to the systemic velocity (615 \kms) of NGC 1023 \citep{Capaccioli+86}, such velocity ranges from -210 $\kms$  to 356 $\kms$. Based on the velocity range, we constructed an \HI column-density map of NGC 1023, as shown in Fig.~\ref{fig:NGC1023_all}\textcolor{blue}{a}. The column-density error on the map is 1$\sigma$=1.3$\times$10$^{18}$ cm$^{-2}$, which is about one order of magnitude lower than that from the WSRT observation for NGC 1023 \citep{Sancisi+84,Morganti+06}. Our high-sensitivity observations of the \HI gas emission in NGC 1023 show a very extended and unbroken structure. NGC 1023 is almost in the center of the extended structure.  In Fig. \ref{fig:NGC1023_all}\textcolor{blue}{b}, the \HI velocity map of the extended structure shows a clear gradient from southwest to northeast. We infer that the extended structure is a rotating gas disk of NGC 1023. As depicted by a dotted line in  Fig. \ref{fig:NGC1023_all}\textcolor{blue}{b}, the gas disk is twisty. And the rotation axis of the gas disk is misaligned with the photometric axis, but it almost matches the bar direction of NGC 1023 \citep{Debattista+02,Bettoni+12}.

The gas disk of NGC 1023 in Fig.~\ref{fig:NGC1023_all}\textcolor{blue}{a} also exhibits an asymmetric outskirts, involving two striking tails (tail NE and tail SE). The tail SE was detected for the first time in NGC 1023 based on our higher-sensitivity \HI observations.  Since no nearby massive galaxies were detected in the direction of the tail SE, this tail cannot be a tidal tail that was dragged out by a galaxy outside NGC 1023. Both the velocity maps of the tail NE and tail SE in Fig. \ref{fig:NGC1023_moment1} shows a gradient towards the center of NGC 1023.  Moreover, the asymmetric outskirts of NGC 1023 connect two dwarf galaxies (NGC 1023C and NGC 1023D), whereas dwarf galaxies NGC 1023A and NGC 1023B appear to be embedded in the gas disk of NGC 1023. Fig. \ref{fig:NGC1023_group} shows the \HI column-density map of each dwarf galaxy. We found that NGC 1023B, NGC 1023C, and NGC 1023D respectively connected with a part of the gas disk of NGC 1023 by a bridge.  The simulations show that when two galaxies interact, they will result in tidal tails in each galaxy  \citep{Toomre+72,Springel+05}. In particular, they will form a connecting bridge, which could provide the most direct evidence that two galaxies are interacting.
The bending of the tail NE towards NGC 1023C indicates that this tail may have formed because of a tidal effect from this dwarf galaxy. We estimated that the ratio ($M_{\rm dyn}$/$M_{\rm bar}$) of  the dynamical masses to the baryons masses is $\sim$11.5 for NGC 1023C (See Appendix B). From previous observations and numerical simulations, $M_{\rm dyn}$/$M_{\rm bar}\simeq$1 is a characteristic of tidal dwarf galaxies  \citep{Lelli+15,Duc+04,Bournaud+06}.  Both the NGC 1023B and NGC 1023D are similar in kinematics to the NGC 1023C. We suggest that these three dwarf galaxies are interacting with NGC1023, not the tidal dwarf galaxies that formed when galaxies merged.

\begin{figure}rifield, 
\centering
\includegraphics[width=\textwidth]{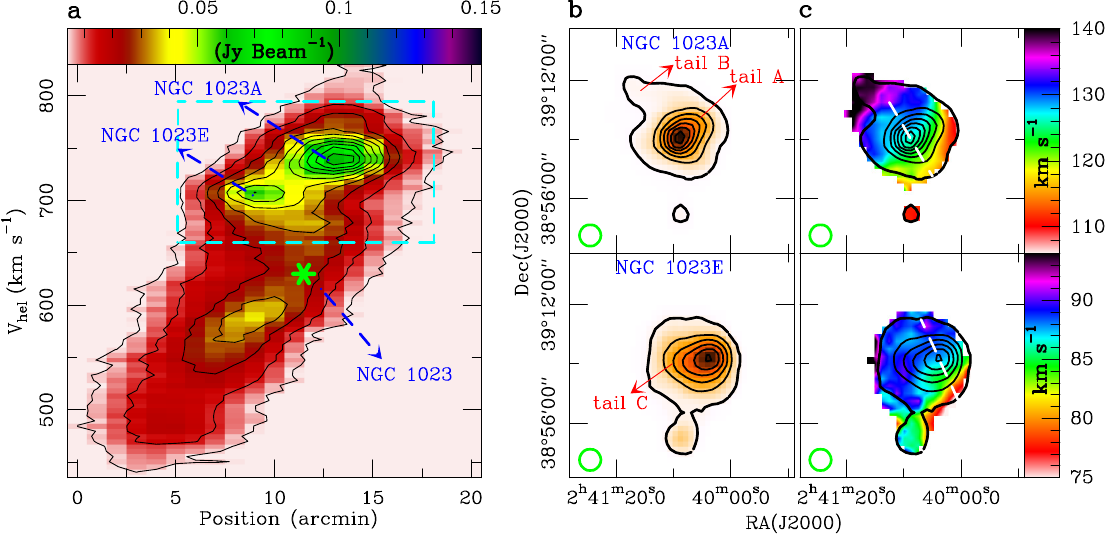}
\caption{{\noindent A galaxy pair within the gas disk of NGC 1023}. {\bf a},  position-velocity (PV) diagram in colour scale overlaid with black contours. The black contours begin at 3$\sigma$ (1$\sigma$=1.3 mJy beam$^{-1}$) in steps of 7$\sigma$. We use a cyan dashed box to mark the galaxy pair in the PV diagram.  {\bf b} and  {\bf c}, \HI column-density and velocity-field maps. The black contours begin at 3.0$\times$10$^{18}$ cm$^{-2}$ in steps of 1.8$\times$10$^{19}$ cm$^{-2}$ in each panel. The while dashed line in each  velocity-field map marks the direction of the velocity gradient. These velocities are relative to the systemic velocities  of NGC 1023A and NGC 1023E, respectively. The FAST beam in green circle is shown in the bottom-left corner of each panel. We adopted the integrated-velocity range from 108.2 $\kms$ to 147.5  $\kms$ for NGC 1023A, while from 73.7 $\kms$ to 108.2  $\kms$ for NGC 1023E.}
\label{fig:NGC1023_FAST}
\end{figure}

To analyze the kinematics of NGC 1023 in velocity space, we created velocity and position (PV) diagrams through its center with a half-beam width and a length of 21$^{\prime}$. The cutting directions start from the north and run counterclockwise every 15$^{\circ}$. From the PV diagrams in Fig. \ref{fig:NGC1023_PV}, we found that NGC 1023 shows a  rotational gas disk that hosts several gas clumps. In the highest column-density region of the gas disk, we identified a pair of gas clumps, as indicated by the blue boxes. To illustrate the identified pair system, we made a special PV diagram with cutting direction along the connecting axis of the pair members. The identified pair are highlighted by a cyan dashed box in Fig.~\ref{fig:NGC1023_FAST}\textcolor{blue}{a}. The spectroscopic observations indicate that the heliocentric radial velocity of NGC 1023 is  615$\pm$20 \kms, whereas that of NGC 1023A is 742$\pm$30 \kms. By comparing these velocities to the PV diagram in Fig.~\ref{fig:NGC1023_FAST}\textcolor{blue}{a}, we found that one member of the pair is associated with NGC 1023A, and the other is a newly discovered \HI clump, named at NGC 1023E. We used the Source Finding Application (SoFiA2) pipeline \citep{Westmeier+21} to determine the radial velocity and coordinate position of NGC 1023E. By comparing these systemic velocities, NGC 1023E with a radial velocity of 704.2$\pm$1.6 \kms seems to be located between NGC 1023  and NGC 1023A in the line of sight.

\begin{figure}
\centering
\includegraphics[width=0.85\textwidth]{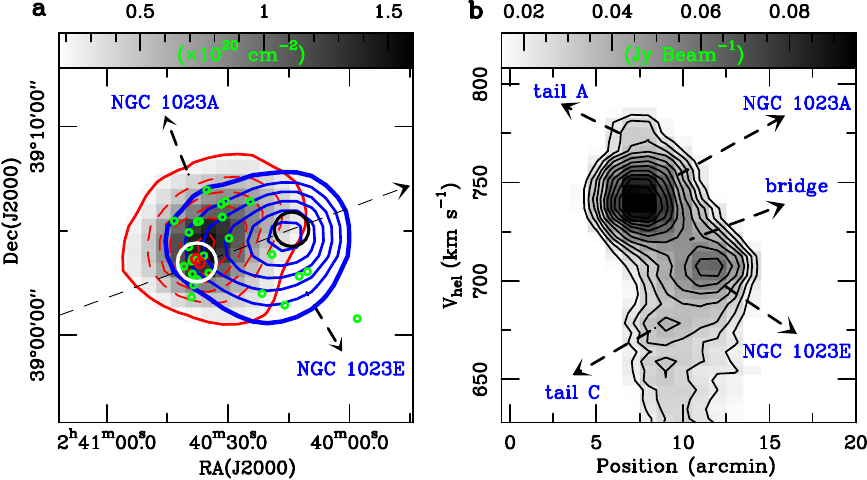}
\caption{Morphology of the galaxy pair.  {\bf a}, the \HI column-density maps of the dense regions of NGC 1023A (red contours and black scale) and NGC 1023E (blue contours). Both blue and red contours begin at 2.0$\times$10$^{19}$ cm$^{-2}$. The white circle represents the predicted optical radius of NGC 1023A, while the black circle represents the predicted optical radius of NGC 1023E. Previously identified blue Faint fuzzies (FFs) \citep{Forbes+14} and two stellar clusters \citep{Larsen+02} are indicated by the green and red circles, respectively. {\bf b}, PV diagram in black scale overlaid with black contours. The cutting direction is shown in a black arrow in panel a. The black contours begin at 20$\sigma$ (1$\sigma$=1.3 mJy beam$^{-1}$) in steps of 4$\sigma$.
}
\label{fig:NGC1023_FA}
\end{figure}

NGC 1023A and NGC 1023E in Fig. \ref{fig:NGC1023_FAST}\textcolor{blue}{a} are connected not only in space but also in velocity, implying that there is an \HI gaseous bridge between them. Moreover, both the dense gases within NGC 1023A and NGC 1023E show a bowed head-tail structure in the column-density maps of Fig. \ref{fig:NGC1023_FAST}\textcolor{blue}{b}.  Fig. \ref{fig:NGC1023_FA}\textcolor{blue}{a} shows the overlapping maps of the dense regions of NGC 1023A and NGC 1023E. We found that the orientation of these identified tails is not random. The tail of NGC 1023A bends towards the head of NGC 1023E. Similarly, the tail of NGC 1023E bends towards the head of NGC 1023A. In order to confirm whether these are true tidal tails, we also made an PV diagram with the cutting direction shown in Fig. \ref{fig:NGC1023_FA}\textcolor{blue}{a}. The PV diagram in Figs. \ref{fig:NGC1023_FA}\textcolor{blue}{b} make the bowed head-tail structures in  NGC 1023A and NGC 1023E more obvious. And both tails are high speed relative to their central dense cores. It closely resembles the traits of the tidal tail. The bridge and tidal tails detected in the pair system imply that NGC 1023A and NGC 1023E are interacting. Furthermore, as shown in Fig \ref{fig:NGC1023_FA}\textcolor{blue}{a}, previously identified blue Faint fuzzies (FFs) with a few 100 Myr coincide well with the interacting system in spatial distribution \citep{Forbes+14}. The FFs are a relatively new class of star cluster.  Besides, two blue clusters with ages between 125 and 500 Myr are identified in NGC 1023A on the basis of their strong Balmer lines \citep{Larsen+02}. The enhanced star formation activity in NGC 1023A further confirmed that NGC 1023A is interacting with NGC 1023. 

The formation of a pair system with the dwarf galaxy NGC 1023A suggests that NGC 1023E is probably not a simple gas clump. In  Fig.~\ref{fig:NGC1023_FAST}\textcolor{blue}{c}, the moment 1 maps of both NGC 1023A and NGC 1023E show a relatively regular velocity gradient, which is regarded to be a key characteristic of galaxies. In order to verify the existence of the velocity gradients in NGC 1023A and NGC 1023E, we made the new PV diagrams along their velocity gradients, denoted by the white dashed lines visible in Fig.~\ref{fig:NGC1023_FAST}\textcolor{blue}{c}. The PV diagrams in Fig. \ref{fig:NGC1023_rot}  confirm that both NGC 1023A and NGC 1023E have a velocity gradient.  Additionally, we estimated the \HI gas masses ($M_{\rm HI}$) of NGC 1023A and NGC 1023E to be 2.7$_{-0.3}^{+0.3}\times10^{8}$ \msol and 2.0$_{-0.2}^{+0.2}\times10^{8}~\msol$, and their effective radius to be 20.0$\pm$0.4 kpc and 18.7$\pm$0.4 kpc, respectively, indicating that NGC 1023E has a similar size and \HI masses with NGC 1023A. Besides, both the $M_{\rm dyn}$/$M_{\rm bar}$ are $\sim$5.5, which indicates that the dark matter dominates in NGC 1023A and NGC 1023E. All of these evidences suggest that NGC 1023E can be classified as a galaxy, like NGC 1023A.

\begin{figure}
\centering, 
\includegraphics[width=0.9\textwidth]{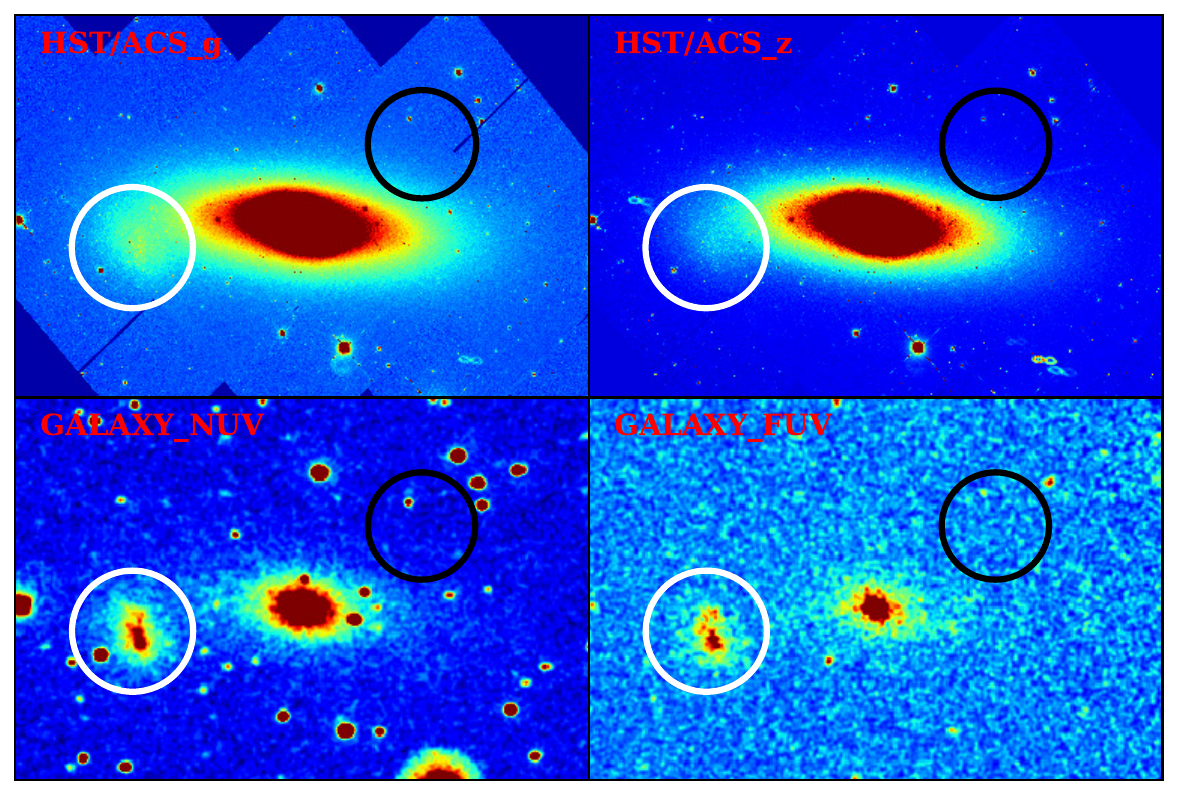}
\caption{Optical and ultraviolet (UV) emission maps of the galaxy pair members. The Hubble Space Telescope (HST) $g$ and $z$ bands data are used to trace the optical emission, and the GALEX NUV and FUV data for the UV emission. Fluxes in each band are on an arbitrary unit. The white circle represents the predicted optical radius of NGC 1023A, while the black circle for the predicted optical radius of NGC 1023E.}
\label{fig:NGC1023-optical}
\end{figure}

In order to investigate the presence of stellar components that could be associated with NGC 1023E, we used Hubble Space Telescope (HST) g- and z-bands, and GALEX NUV and FUV images. Previous studies established a strong correlation between the \HI mass ($M_{\rm HI}$) and the optical disc diameter ($D_{25}$) in kpc for late-type galaxies, as shown by the equation log$M_{\rm HI}$=a+b$\times$log$D_{25}$, where the coefficients could be adopted by $a$=7.0 and $b$=1.95 \citep{Broeils+97,Toribio+11}.  We can make the restriction on the scales of potential optical emission from NGC 1023E using the correlation. Finally, the predicted values for $D_{25}$ are 5.4$\pm$0.6 kpc for NGC 1023A, and 4.7$\pm$0.5 kpc for NGC 1023E. As seen in Fig. \ref{fig:NGC1023-optical}, the optical emission of NGC 1023A is just within the predicted optical scale. But for NGC 1023E, we did not detect any extended optical or ultraviolet (UV) emission in the predicted radius of NGC 1023E, except for a foreground star with a parallax of 0.76$\pm$0.08 mas, which is identified by using the Gaia star diagram \citep{Gaia+16}. 

\section{Discussions and Summary}

Neither the fading spirals nor the result of galaxy mergers can explain the stellar kinematics of NGC 1023, implying that it is necessary to explore other mechanisms for formation of this S0 galaxy. From our high-sensitive \HI observations, NGC 1023 clearly displays an extended warped gas disk.  We obtained that the total mass of the gas disk is about $2.0\times10^{9}  \msol$ (See Appendix A). It was previously thought that the irregular \HI gas detected in NGC 1023 was stripped gas from its companion NGC 1023A \citep{Cortesi+16,Dolfi+21}. We observed that NGC 1023A is still a gas-rich dwarf galaxy (See Fig. \ref{fig:NGC1023_FAST}\textcolor{blue}{b}), and thus excluded the possibility that the \HI gas in NGC 1023 is donated by this companion. In addition, \HI observations of galaxies in the local Universe have indicated that extended warped disks are usually related to spiral galaxies \citep{Briggs+90,Garcia-Ruiz+02}. The warps are thought to originate from interactions with nearby galaxies and cold-mode accretion \citep{Sancisi+08,van+11,Rahmani+18}. In this study, we found that NGC 1023 is interacting with four dwarf galaxies. The largest galaxy NGC 1023A among them seems to have fallen into the gas disk of NGC 1023, forming a bright-dark galaxy pair. This strongly supports that the warped gas disk of NGC 1023 originated from  interactions. The warped gas disk could be a fossil record of the progenitor galaxy of NGC 1023. This result may favour a model in which NGC 1023 formed from a spiral by tidal interactions. 

Tidal interaction is also considered as a possible physical mechanism for the transformation of spirals into S0 galaxies in galaxy groups from the chemodynamical simulations \citep{Bekki+11}. The simulated S0 galaxies transformed from spirals in this mechanism have young and metal-rich stellar populations in their bulges. The simulated S0 galaxies have a lower rotation amplitude ($V_{\rm rot}/\sigma\sim1$) and a flatter radial velocity-dispersion profiles. The cold \HI gas adjacent to the simulated S0 galaxies presents unique spatial distributions such as rings, long tails, and massive \HI clouds without optical counterparts.  NGC 1023 has a nuclear disc, which is characterized by a younger ($\sim$3.4 Gyr) and more metal-rich ([Fe/H]= 0.50 dex) stellar population in its bulge \citep{Corsini+16}. The stellar and GCs/PNe kinematics reveal that  NGC 1023 is more pressure supported at its large optical radii ($V_{\rm rot}/\sigma<1$), and has the relatively flatter profiles in its radial velocity dispersion \citep{Noordermeer+08,Dolfi+21}. Especially, the gas disk of NGC 1023 has two long tails (tail NE and tail SE) and an \HI  gas clump without an optical counterpart (NGC 1023E). The same chemical, kinematical and structural properties in NGC 1023 as in the simulated S0s suggest that NGC 1023 is formed via the morphological transformation from a spiral under the tidal interactions. However, there is a difference that NGC 1023 is a gas-rich S0 galaxy, while the simulated S0 galaxies end up being gas poor. This may be due to that the simulated S0 galaxies are interacting with other similarly massive or more massive galaxies within the group \citep{Bekki+11}, while in NGC 1023 the interactions are with a number of dwarf galaxies. The relatively strong interactions with massive galaxies may strip away gas from the predecessor of S0 galaxy.

The previous model predicated that the majority of gas in galaxies with baryonic masses more than $10^{9}$ \msol will become Toomre unstable and result inevitably in star formation \citep{Taylor+05}. The formed NGC 1023 is a gas-rich S0 galaxy with an \HI mass of $2.0\times10^{9}  \msol$. We also obtained that the peak column density is $\sim3.3\times10^{20}$ cm$^{-2}$ in its gas disk of NGC 1023, which is slightly less than that of 5.0$\times10^{20}$ cm$^{-2}$ from the WSRT observation. The peak value is near the column-density threshold (5.0$\times10^{20}$ cm$^{-2}$) for star formation \citep{Gallagher+84,Davies+06}. This suggests that although there is a significant amount of cold gas in NGC 1023, the forming star in the cold gas has been suppressed. The quenching of the forming star in NGC 1023 is not due to a lack of gas, but that is likely to be caused by the external tidal effect.

For the formation of S0 galaxies due to tidal-interaction mechanism, it can create large \HI gas clumps from the gas stripped from spirals being transformed into the S0 galaxies \citep{Bekki+11}. If the clump is self-gravitating, it could become a small dwarf galaxy that also retains little dark matter from their parents' haloes \citep{Barnes+92}. We also detected a large gas clump NGC 1023E with an \HI mass of approximately 2.0$\times10^{8} \msol$ in the gas disk of NGC 1023.  NGC 1023E shares the same morphology and dynamics as dwarf galaxy NGC 1023A. Furthermore, it can be observed that the \HI gas present in NGC 1023E can be simulated as a rotating gas disk (as outlined in Appendix B). Therefore, NGC 1023E has all the characteristics of a galaxy but no optical counterpart from the deep HST observations. The  dominant presence of dark matter ($M_{\rm dyn}$/$M_{\rm bar}\simeq$5) in  NGC 1023E suggests that it is  likely to be a newly formed starless galaxy, rather than a tidal dwarf galaxy, which is generally loctated in tidal tails and are absence of the dark  matter ($M_{\rm dyn}$/$M_{\rm bar}\simeq$1) \citep{Lelli+15}. Based on the HST $g$-band image, we estimated that the upper limit surface brightness of NGC 1023E is $\mu_{g}$ = 33.8 mag arcsec$^{-2}$.

In sum,  for the first time in observations,  we have confirmed  that a massive S0 galaxy in a galaxy group can be generated through the joint action of multiple tidal interactions.  The fact that about half of S0 galaxies reside in groups \citep{Crook+07,Wilman+09}. It is possible that tidal interaction is an important mechanism for the formation of S0 galaxies. Future observational studies with bigger sets of samples should reveal a clearer picture of S0 galaxies in groups.

\acknowledgments 
We thank the referee for insightful comments that improved the clarity of this manuscript. We acknowledge the supports of the National Key R$\&$D Program of China No. 2022YFA1602901. This work is also supported by the National Natural Science Foundation of China (Grants No. 12373001 and 11933011), and the Central Government Funds for Local Scientific and Technological Development (No. XZ202201YD0020C).

\begin{table}[h]
\footnotesize
\centering
\caption{\small Measured and derived properties of the dwarf galaxies. We list: Equatorial coordinates (RA, Decl);  system velocity ($V_{\rm sys}$);  inclination angle ($i$); rotational velocity ($V_{\rm rot}$); velocity dispersion ($\sigma_{\rm v}$); effective radius ($R_{\rm eff}$); absolute $g$-band and $r$-band magnitudes ($m_{g}$, $m_{r}$); \HI gas mass ($M_{\rm \hi}$); steller mass ($M_{\rm \star}$);  dynamic mass ($M_{\rm dyn}$); $R$ is the ratio of  the dynamical masses to the baryons masses.}
\label{tab:prop}
\setlength{\tabcolsep}{4.2pt}
\begin{tabular}{lccccccccccccc}
\noalign{\vspace{5pt}}\hline\hline\noalign{\vspace{5pt}}
 Name & RA & Decl & $V_{\rm sys}$ &$i$&$V_{\rm rot}$&$\sigma_{\rm v}$&$R_{\rm eff}$&$m_{g}$&$m_{r}$ & log$M_{\rm HI}$ &log$M_{\rm \star}$ & log$M_{\rm dyn}$ & $R$ \\
 NGC & [$\circ$] &[$\circ$]&[km/s]&[$\circ$]&[km/s]&[km/s]& [kpc] &[mag]&[mag]&[$\msol$]&[$\msol$]&[$\msol$]\\
\noalign{\vspace{5pt}}\hline\hline\noalign{\vspace{5pt}}
1023A & 40.1571 & 39.0575 &  740.7$^{+0.5}_{-0.5}$ & 43$^{+2}_{-2}$& 17.0$^{+3.2}_{-3.2}$&9.3$^{+1.5}_{-1.5}$ &20.0$^{+1.2}_{-1.2}$ & 14.29 & 14.15 &8.4$^{+0.1}_{-0.1}$ &8.1$^{+0.1}_{-0.1}$ &9.4$^{+0.1}_{-0.1}$ &5.6$^{+0.4}_{-0.4}$  \\
1023C & 40.1650 & 39.3797 &  904.6$^{+1.6}_{-1.6}$ & 81$^{+5}_{-5}$& 5.1$^{+2.9}_{-2.9}$&11.2$^{+1.6}_{-1.6}$ & 12.3$^{+1.3}_{-1.3}$& 17.02 & 16.59 &7.8$^{+0.1}_{-0.1}$&7.2$^{+0.1}_{-0.1}$&9.1$^{+0.1}_{-0.1}$ &11.5$^{+1.0}_{-1.0}$ \\
1023E & 40.0667 & 39.0845 &  704.2$^{+0.7}_{-0.7}$ & 47$^{+3}_{-3}$& 12.9$^{+3.1}_{-3.1}$&7.5$^{+1.4}_{-1.4}$ &18.7$^{+1.1}_{-1.1}$& -- & -- & 8.3$^{+0.1}_{-0.1}$&--&9.2$^{+0.1}_{-0.1}$  &5.5$^{+0.6}_{-0.6}$\\ 
1023B & 40.2500 & 39.0719 &  573.3$^{+2.6}_{-2.6}$ & --  & -- &-- & --& 16.74 & 16.25 & 6.8$^{+0.2}_{-0.2}$ &7.4$^{+0.1}_{-0.1}$ &-- &--\\ 
1023D & 40.1375 & 38.9003 &  699.3$^{+1.6}_{-1.6}$ & -- & -- & -- & -- & 16.62& 15.97 & 7.5$^{+0.1}_{-0.1}$& 7.5$^{+0.1}_{-0.1}$& -- &-- \\
\noalign{\vspace{5pt}}\hline\hline\noalign{\vspace{5pt}}
\end{tabular}
\end{table}

\appendix

\section{\HI column density and baryonic mass of the galaxy.}
\label{sec:Column}
Under the assumption that the gas in galaxies is optically thin for the \HI line, the column density $N(x,y)$ in each pixel can be calculated by $N(x,y)= 1.82\times10^{18}\int T_{\rm B}dv$,  where $dv$ is the velocity width in \kms. We made the column-density distribution maps on the obtained moment 0 map (total intensity) using the measured relevant main beam gain $T_{\rm B}/\it S_{v}$.

The majority of gas in galaxies is \HI and helium. Assuming the same helium-to-\HI ratio as that derived from the Big Bang nucleosynthesis, a factor of 1.33 is included to account for contribution of Helium \citep{Planck+20}.  The baryonic
masses ($M_{\rm bar}$) can be determined by $M_{\rm bar}=M_{\star}+M_{\rm gas}$, where $M_{\star}$ is the obtained total
stellar mass. The total gas mass of galaxies is derived by $M_{\rm gas} = 1.33\times M_{\rm HI}$, where  $M_{\rm HI}$ is \HI gas mass, which can be determined by:
\begin{equation}
    M_\mathrm{HI} =  2.36\times10^{5}D^{2}{\int}S_{v}(x,y)dxdy
\end{equation}
\noindent where  $D$ is the adopted distance  to each galaxy and $S_{v}(x,y)$ is the integrated \HI flux in Jy km s$^{-1}$ in each pixel. Here we take the distance of NGC 1023 as that of each galaxy in this group.   

In order to estimate the total stellar mass of each galaxy, we used the Pan-STARRS1 $g$-band and $r$-band data. The Pan-STARRS1 survey is a 3$\pi$ steradian survey with a medium depth in 5 bands \citep{Chambers+16}. For dwarf galaxies, the stellar mass can be given by \citep{Zhang+17}: 
\begin{equation}
    log(M_{\star}/L_{g})=-0.745+1.616(m_{g}-m_{r})
\end{equation}
where $L_{g}$ is stellar luminosity in $g$-band, which can be determined by $L_{g}$=$D^{2}10^{10-0.4(m_{g}-M_{g}^{0})}$, where $M_{g}^{0}$ is the absolute solar g-band  magnitude, which is adopted as 5.03 mag \citep{Willmer+18}.  In order to measure  the apparent magnitude ($m_{g}$, $m_{r}$), we  made a new sky-background subtraction of each galaxy using the software SEXtractor \citep{Bertin+96} and a row-by-row and column-by-column method \citep{Wu+02,Du+15}. For the starless galaxy NGC 1023E, we can ignore the contribution of stellar mass to the baryonic mass.

\section{Dynamic masses of the galaxy.}
\begin{figure*}
\centering
\includegraphics[width=16.0cm]{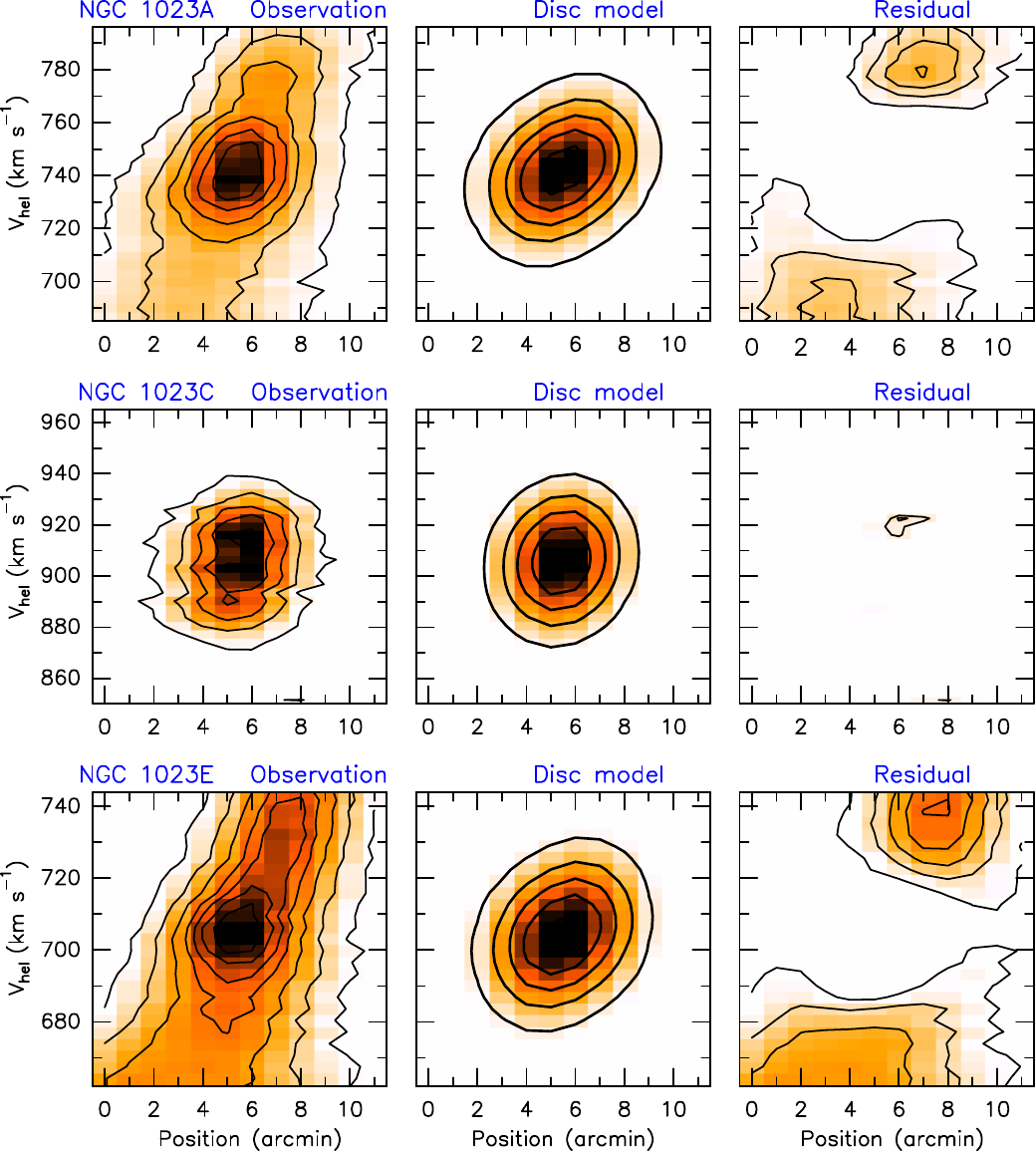}
\vspace{-8pt}
\caption{PV diagrams in colour scale overlaid with the black contours, obtained from observed cube (left), model cube (middle), and residual cube (right). The cutting is through the center of each galaxy along their position angles. All the black contours begin at 5$\sigma$ in each panel. 1$\sigma$=1.3 mJy beam$^{-1}$.
\label{fig:NGC1023_rot}
}
\end{figure*}

In order to obtain the dynamic mass of galaxy, we use the Tilted Ring Fitting Code (TiRiFiC) software package to fit its \HI cube data. The TiRiFiC is a 3D tilted-ring fitting code that has demonstrated considerable success in describing the kinematics and morphology of rotating disks \citep{Jozsa+07}. The Bootstrap method is used to estimate the final values and errors of model parameters. Once we have reached a preliminary final model, we move the individual nodes of the model several times with random values, and then restart the fitting process. We calculated the mean and standard deviation of each model parameter  after 20 such fitting processes and assume that the mean is the final model parameter and its error. Apart from the obtained fitting parameters, we also derived model cube data. With the cube data, we created  PV diagrams of each galaxy in Fig. \ref{fig:NGC1023_rot}, which can be used to compare with observation data and then determine whether the TiRiFiC model fitting is the best fit. After subtracting the disk model, the residuals are below 3$\sigma$ significance in Fig. \ref{fig:NGC1023_rot}, where $\sigma$ is the noise level, suggesting that the majority of the \HI gas emission for each galaxy can be modelled as a rotating disk.

Located adjacent to NGC 1023, there are five other galaxies. Since the \HI emission size of NGC 1023B and NGC 1023D are smaller ($\leq$2 beams), we only successfully fitted other three relatively large dwarf galaxies. The obtained rotational velocities ($V_{\rm rot}$) and velocity dispersion ($\sigma_{\rm v}$) from the fittings are listed in Extended Data Table~\ref{tab:prop}. Because the rotational  velocities are comparable with the velocity dispersion for  NGC 1023A, NGC 1023C, and NGC 1023E, it needs to consider the contribution of velocity dispersion to dynamic mass \citep{Hoffman+96,Roman+21}. We can eatimate dynamic mass by $M_\mathrm{dyn}=(V^{2}_{\rm rot}+3\sigma^{2}_{\rm v})R_{\rm HI}/G$, where  $R_{\rm HI}$ is effective radius and $G$ is gravitational constant. To correct for beam smearing effect, the effective radius can be determined by $R_{\rm HI} = \sqrt{D_{\rm HI}^{2}-B_{\rm F}^{2}}$/2, where $D_{\rm HI}$ is the uncorrected \HI sizes of galaxies, and $B_{\rm F}$  is the FAST beam size (2.9$^{\prime}$). We measured the $D_{\rm HI}$ from their \HI column-density maps. Observed properties and estimated masses are summarized in Extended Data Table~\ref{tab:prop}.

\begin{figure*}
\centering
\includegraphics[width=12cm]{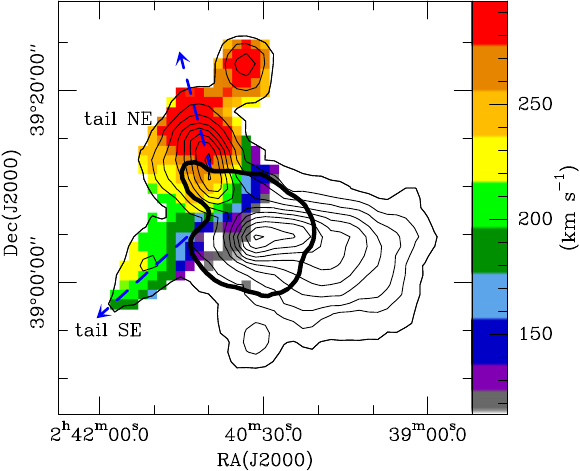}
\vspace{0pt}
\caption{\HI velocity-field map in colour scale overlaid with the \HI column density in the black contours. The velocity ranges from 115 \kms to 295 \kms.  These velocities are relative to the systemic velocity (615 \kms) of NGC 1023. The thick black line  indicates the position and size of NGC 1023A, while the two blue arrows mark the velocity gradients of tail NE and tail SE.
\label{fig:NGC1023_moment1}
}
\end{figure*}

\begin{figure*}
\centering
\includegraphics[width=16cm]{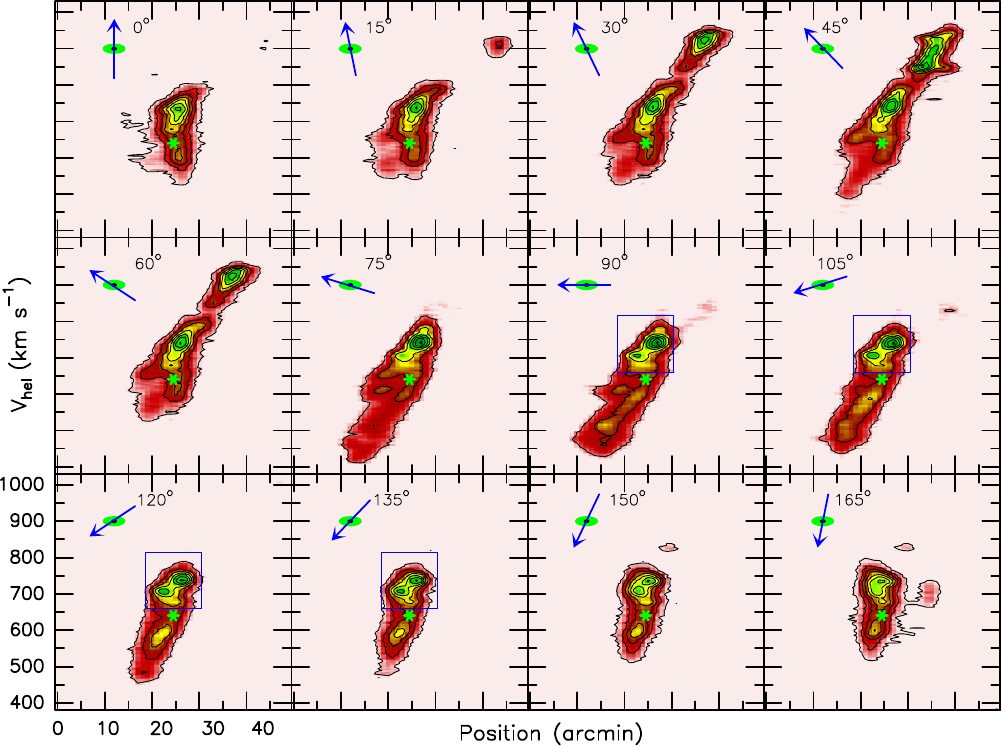}
\vspace{0pt}
\caption{Position-Velocity (PV) diagrams, which cut through NGC 1023 along different position angles, shown in a blue arrows on the top-left corners of each panel.  The blue boxes indicate the binary galaxy, and the green crosses are used to mark the position of S0 galaxy NGC 1023 in the PV diagram.
\label{fig:NGC1023_PV}
}
\end{figure*}

\end{document}